# Feeder Load Balancing using Fuzzy Logic and Combinatorial Optimization-based Implementation

Abhisek Ukil

*(Corresponding Author)*

*ABB Corporate Research, Baden-Daettwil, Switzerland*

*Address: Segelhofstrasse 1K, CH-5405, Baden 5 Daettwil, Switzerland*

*Tel: +41 58 586 7034*

*Fax: +41 58 586 7358*

*E-mail:* abhiukil@yahoo.com

Willy Siti

*Tshwane University of Technology, Pretoria, South Africa*

*E-mail:* willysiti@yahoo.com

**Abstract**

The distribution system problems, such as planning, loss minimization, and energy restoration, usually involve the phase balancing or network reconfiguration procedures. The determination of an optimal phase balance is, in general, a combinatorial optimization problem. This paper proposes a novel reconfiguration of the phase balancing using the fuzzy logic and the combinatorial optimization-based implementation step back to back. Input to the fuzzy step is the total load per phase of the feeders. Output of the fuzzy step is the load change values, negative value for load releasing and positive value for load receiving. The output of the fuzzy step is the input to the load changing system. The load changing system uses combinatorial optimization techniques to translate the change values (kW) into number of load points and then selects the specific load points. It also performs the inter-changing of the load points between the releasing and the receiving phases in an optimal fashion. Application results using the distribution feeder network of South Africa are presented in this paper.

# 1    Introduction

The distribution system technology has changed drastically, both qualitatively and quantitatively. With the increasing technological development, the dependence on electric power supply has increased considerably. While demand has increased, the need for a steady power supply with minimum power interruptions and fast fault restoration has also increased. To meet these demands, automation of the power distribution system needs to be widely adopted.  All switches and circuit-breakers involved in the controlled networks are equipped with facilities for remote operation. The control interface equipments must withstand extreme climatic conditions. Also, control equipments at each location must have a dependable power source.  To cope with the complexity of the distribution, the latest computer, communication and distribution technologies are needed to be employed. The distribution automation can be defined as an integrated system concept. It includes control, monitoring and some times, decision to alter any kind of loads. The automatic distribution system provides directions for automatic reclosing of the switches and remote monitoring of the loads contributing towards phase balancing.

Phase balancing is very important and usable operation to reduce distribution feeder losses and improve system security. There are a number of normally closed and normally opened switches in a distribution system. By changing the open/close status of the feeder switches, load currents can be transferred from feeder-to-feeder, i.e. from heavily loaded to less loaded feeders. In South Africa, to reduce the unbalance current in a feeder the connection phases of some feeders are changed manually after some field measurement and software analysis. Although in some cases this process can improve the phase current unbalance, this strategy is more time-consuming and erroneous. In this paper, we propose the use of fuzzy logic-based load balancing and combinatorial optimization-based changing system for load change implementation as novel procedures to perform the feeder load balancing.

The remainder of this paper is organized as follows. In section-2, the problem of feeder load (phase) balancing is explained briefly. In section-3, we describe the specific load balancing problem addressed in this paper. Fuzzy logic-based load balancing technique is presented in details along with application examples in section-4. Section-5 discusses about the combinatorial optimization-based system architecture for implementing the load changes, along with application example. Assumptions, generalizations and discussions of the load balancing techniques considered in this paper are presented in section-6, and conclusions are given in section-7.

## 2  Phase Balancing

In South Africa, to reduce the degree of the phase current unbalance, thus avoiding the malfunctioning of the protective relay and unintentional service discontinuity, the connection phases of some critical distribution transformers are usually changed manually following many field measurements and analysis. In some cases, this process certainly improves the phase voltage and currents unbalances. However, considerable time must be spent to achieve an acceptable result. In addition, the balancing status of the system, most of the time, lasts only for a short time, sometimes even only an hour. This consequence is expected because the time varying characteristic of the load is usually not considered in detail in the trial and error approach.

In most of the cases, the phase voltage and current unbalances can be greatly improved by suitably arranging the connection phases between the distribution transformers and a primary feeder. It is also possible to advance the phase current unbalances in every feeder segment by means of changing the connection phases [1]. The phase voltage unbalances along a feeder can also be improved in common cases.

In general, distribution loads show different characteristics according to their corresponding distribution lines and line sections. Therefore, at the load levels, each time period can be regarded as non-identical. In the case of a distribution system, with some overloaded and some lightly loaded branches, there is the need to reconfigure the system such that loads are transferred from heavily loaded to less loaded feeders. Here the maximum load current the feeder conductor can take may be taken as the reference [4]. Nonetheless, the transfer of load must be such that a certain predefined objective is satisfied. In this case, the objective is for the ensuing network to have minimum real power loss. Consequently, phase balancing may be redefined as the rearrangement of the network such as to minimize the total real power losses arising from line branches. Mathematically, the total power loss [4] may be expressed as

$$\sum_{i=1}^{n} r_i \frac{P_i^2 + Q_i^2}{|V_i|^2}, \qquad (1)$$

where $r_i$, $P_i$, $Q_i$, $V_i$ are respectively the resistance, real power, reactive power and voltage of the branch $i$, and $n$ is the total number of branches in the system.

*2.1    State-of-the art*

To improve the system efficiency in the modern power distribution systems, sectionalizing switches and tie switches for feeder reconfiguration are used extensively. Baran and Kelly used the state estimation technique for feeder reconfiguration [3]. Ukil, Siti and Jordaan [4], Siti et al. [5] presented the way to control the tie switches using heuristic combinatorial optimization-based methods. The only disadvantage with the tie-switch control is that, in most of the cases, it makes the current and the voltage unbalances worse [2]. There have been different studies concerning the loss minimization of distribution feeders cited in [7-9].

With the advent of artificial intelligence, telecommunication and power electronics equipments in power systems, it is getting easier to envisage automation of the phase and load balancing problem. The automation implementation will be technically advantageous as well as economical for the utilities and the customers, in terms of the variable costs reduction and better service quality, respectively. Ukil, Siti and Jordaan presented the use of the neural networks to find the optimum switching option of the loads among the different phases in [5]. Walters and Schele [1], Chen and Cherng [10] presented the use of genetic algorithms for improving system unbalance and loss minimization. On the basis of these results, other networks identify the radial topology satisfying the optimal condition.

However, no specific technique or combination of techniques has hitherto been identified as the most effective one. Therefore, in this problem of phase balancing we wanted to explore different techniques to judge the suitability. With works using neural networks having been already reported, we wanted to use fuzzy logic in this problem as a comparison. This is because of considerable interests and instances of using fuzzy logic in power systems [11-12]. Kashem, Jasmon and Ganapathy presented the three phase load balancing in distribution system using index measurement techniques in [13]. It was improved by the reconfiguration system using the fuzzy multi-objective approach [14]. Therefore, we wanted to explore fuzzy logic for phase balancing. Another reason is that if fuzzy logic-based approach is successful this could be easily implemented using FPGA architecture.

## 3    Problem Description

### 3.1    Representation of the Feeder

The distribution feeder is usually a three –phase, four wire system with a radial or open loop structure. To improve the system phase voltage and current unbalances, the connection between the specific feeder and the distribution transformer should be suitably arranged, The loads are connected, as in most cases, in a single-phase. For the problem in this paper, we assume that each feeder contains 50 loads or connections to it, following average load flow studies in South Africa [4]. So, the total load to the three feeders can be 150 connections as shown in Fig. 1. In Fig. 1, each load, through the tie-switches, can be connected only to the one of the three phases.

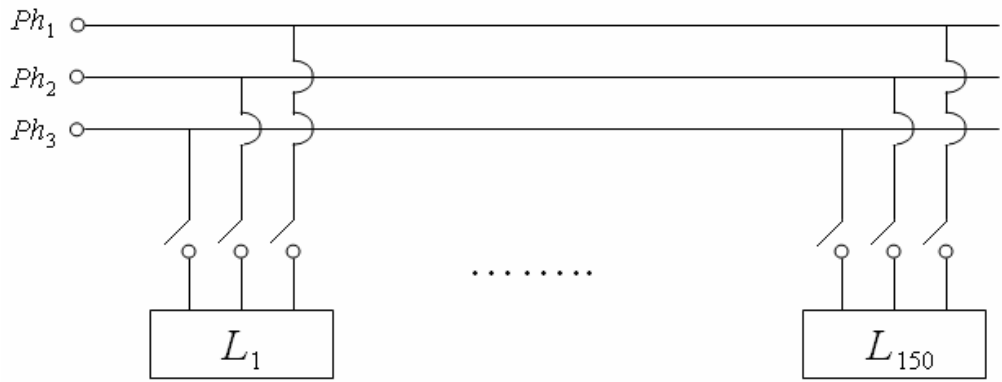

Fig. 1. Example distribution feeder.

*3.2    Current Phase Balancing Technique*

Most of the township houses in South Africa uses the average of three kilowatt power. The major electricity usage is for lighting and domestic works in domestic environment. However, sudden power increase, like the use of heater etc, oftentimes introduces an unknown power in the distribution system, making the transformers to burst and the cables to burn, causing unbalance in the network. To balance the network, the engineers and the technicians must change the phases manually after some field measurement. The changes made to upgrade transformation in different area affect the size of the conductor, but in most of the cases, the size of the phase conductor for the entire line of the feeder is the same.  However, the number of phase conductors may be different in different sections for economic reasons.

The current technique could be improved by changing the open/close status of the feeder switches so that load currents can be transferred from feeder-to-feeder, i.e. from heavily loaded to less loaded feeders. However, it would help if this possibility exists. The motivation of this study is to investigate about the potential of such an approach so that if required it could be implemented for critical load areas like industrial loads etc by incorporating autonomous controller for three-phase change-over possibilities. This would ease out the manual cumbersome works undertaken currently at the expense of phase change-over facility installations.

The power losses depend on the real and the reactive power flows, which are related to the active and reactive loads. Minimum power loss reconfiguration is aimed at by the means of controllable switch-breakers installed at each of the connection on the network feeders, since both the loads and the switch breaker status are physically distributed. In the general formulation of the phase balancing problem, the load values are the independent variables, whereas the switch-breaker statuses are the optimization variables. The objective can be fulfilled performing a control strategy in which the status of each switch breaker depends on the total loads from each feeder. In this way, the network can be optimally operated and it is not necessary to know the load in advance.

For the real implementation of a control system, the following elements are necessary:
- A measurement system for real loads
- A transmission data system for the load data connecting each point
- A transmission system for sending the input signals to the switch breaker.

The control cannot start if the above described components and system are not properly installed and in correct condition.

*3.3    Proposed Technique*

For the above-mentioned system, we propose, in this paper, a fuzzy logic-based load balancing technique along with combinatorial optimization oriented system for implementing the load change decision. The architecture of the proposed system is shown in Fig. 2.

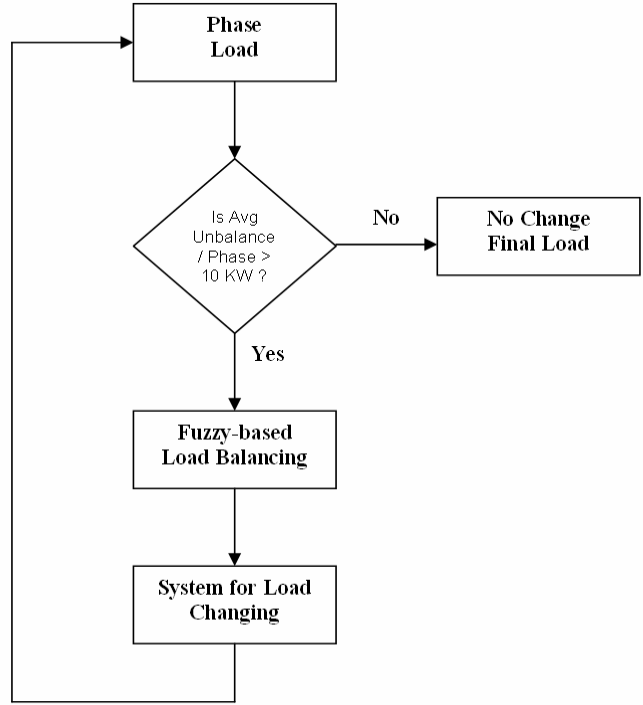

Fig. 2. Architecture for the proposed load balancing system.

$$\text{Avg Unbalance/Ph} = \frac{|Load_{Ph1} - Load_{Ph2}| + |Load_{Ph2} - Load_{Ph3}| + |Load_{Ph3} - Load_{Ph1}|}{3}. \qquad (2)$$

In Fig. 2, the input is the total phase load (for each of the three phases). The average unbalance per phase, calculated according to (2), is checked against a threshold of 10 kW. If the average unbalance per phase is below 10 kW, we can assume that the system is more or less balanced and discard any further load balancing. Otherwise, we go for the fuzzy logic-based load balancing. The output from the fuzzy-based load balancing step is the load change values for each phase. A *negative* value indicates that the specific phase has surplus load and should *release* that amount of load, while a *positive* value indicates that the specific phase is less-loaded and should *receive* that amount of load. This load change configuration is the input to the implementation system which tries to optimally shift the specific number of load points. However, sometimes the implementation system may not be able to execute the exact amount of load change as directed by the fuzzy step. This is because the actual load points for any phase might not result in an optimum combination which sums up to the exact change value indicated by the fuzzy step. So, we implement the best possible change from the implementation system and iteratively check the system unbalance until we achieve the average unbalance below 10 kW, if achievable.

## 4      Fuzzy Logic-based Load Balancing

In this section, we describe the fuzzy logic-based load balancing technique in details. As described in section 3, we assume the average per phase capacity of the system to be 150 kW, with 50 load points connected to any specific phase. For designing the fuzzy controller, we further assume the maximum overload capacity of any phase to be 300 kW. Beyond 300 kW the fuzzy controller should not be used for load balancing. Because, in any case, when any phase reaches its 200% overload condition, it should be cut out from the service to prevent power breakdown and severe overloading of the transformer.

*4.1      Fuzzy Controller: Input and Output*

To design the fuzzy controller, first we design the input and the output variables. For the load balancing purpose, as described in section 3.3, we choose the input as 'Load', i.e., the total phase load (kW) for each of the three phases, and the output as 'Change', i.e., the change of load (kW, positive or negative) to be made for each phase. For the input variable, Table 1 shows the fuzzy nomenclature, and Fig. 3 the respective triangular fuzzy membership functions [15]. And for the output variable, Table 2 shows the fuzzy nomenclature, and Fig. 4 the corresponding triangular fuzzy membership functions [15].

Table 1. Fuzzy nomenclature for the input variable.

| SL. No. | Input (Load) Description | Fuzzy Nomenclature | KW Range |
|---|---|---|---|
| 1 | Very Less Loaded | VLL | 0 to 50 |
| 2 | Less Loaded | LL | 35 to 85 |
| 3 | Medium Less Loaded | MLL | 65 to 115 |
| 4 | Perfectly Loaded | PL | 100 to 150 |
| 5 | Slightly Overloaded | SOL | 125 to 175 |
| 6 | Medium Overloaded | MOL | 165 to 215 |
| 7 | Overloaded | OL | 200 to 250 |
| 8 | Heavily Overloaded | HOL | 235 to 300 |

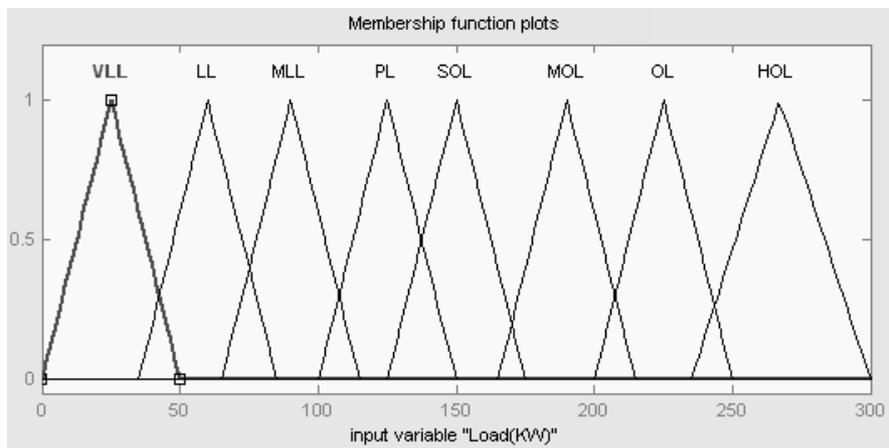

Fig. 3. Fuzzy membership functions for the input variable.

Table 2. Fuzzy nomenclature for the output variable.

| SL. No. | Output (Change) Description | Fuzzy Nomenclature | KW Range |
|---|---|---|---|
| 1 | High Subtraction | HS | -150 to -85 |
| 2 | Subtraction | S | -100 to -50 |
| 3 | Medium Subtraction | MS | -65 to -15 |
| 4 | Slight Subtraction | SS | -50 to 25 |
| 5 | Perfect Addition | PA | 0 to 50 |
| 6 | Medium Addition | MA | 35 to 85 |
| 7 | Large Addition | LA | 65 to 115 |
| 8 | Very Large Addition | VLA | 100 to 150 |

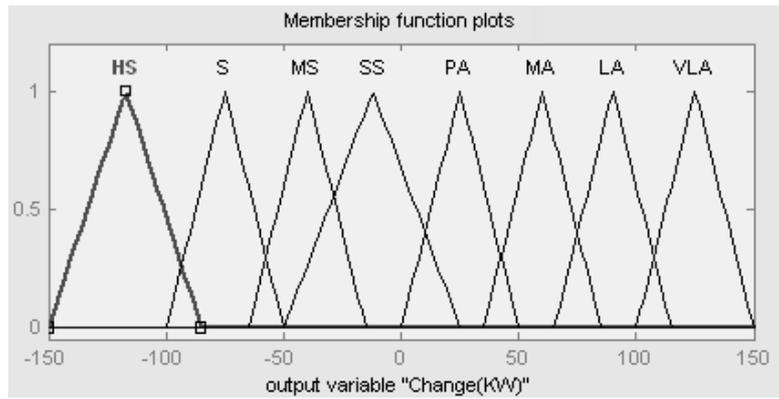

Fig. 4. Fuzzy membership functions for the output variable.

*4.2    Fuzzy Rules and Surface*

Next we determine the IF-THEN fuzzy rule set [15] governing the input and output variable as described in Table 3.

Table 3.  Fuzzy rules for the input and the output variable.

| Rule No. | Rule Description |
|---|---|
| 1 | If *Load* is VLL then *Change* is VLA |
| 2 | If *Load* is LL then *Change* is LA |
| 3 | If *Load* is MLL then *Change* is MA |
| 4 | If *Load* is PL then *Change* is PA |
| 5 | If *Load* is SOL then *Change* is SS |
| 6 | If *Load* is MOL then *Change* is MS |
| 7 | If *Load* is OL then *Change* is S |
| 8 | If *Load* is HOL then *Change* is HS |

Corresponding to the fuzzy input, output variables and the associated rule set, the fuzzy surface [15] is shown in Fig. 5, depicting the nonlinear relationship between the input and the output variable.

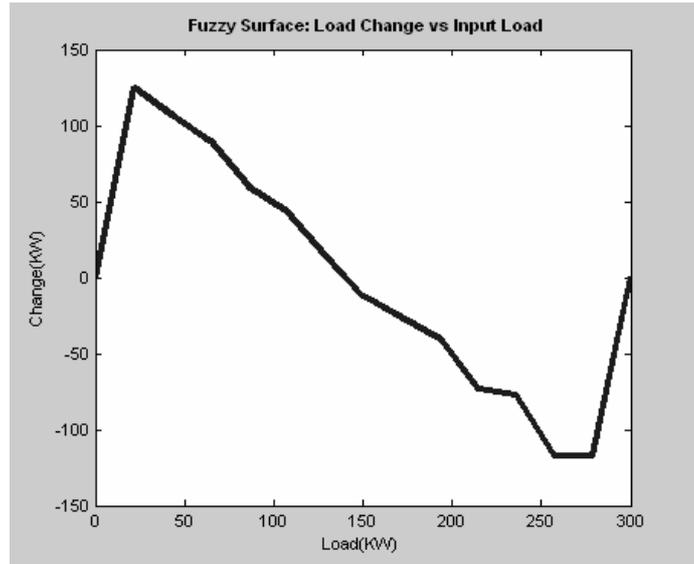

Fig. 5. Nonlinear relationship between the input and the output variable.

*4.3 Application Results*

In this section, we show the application results using the fuzzy logic-based load balancing technique. Matlab® fuzzy toolbox [16] was used for the simulation. We have utilized the Mamdani [17] fuzzy inferencing technique. We take an example input load configuration of [245 120 82] kW for the three phases. We try to balance it using the fuzzy controller described above. The graphical determination of the output load change for the three phases corresponding to this input load and involving the eight fuzzy rules are shown in Fig. 6 to 8. The defuzzification operation is based on the Mamdani (centroid) technique [17].

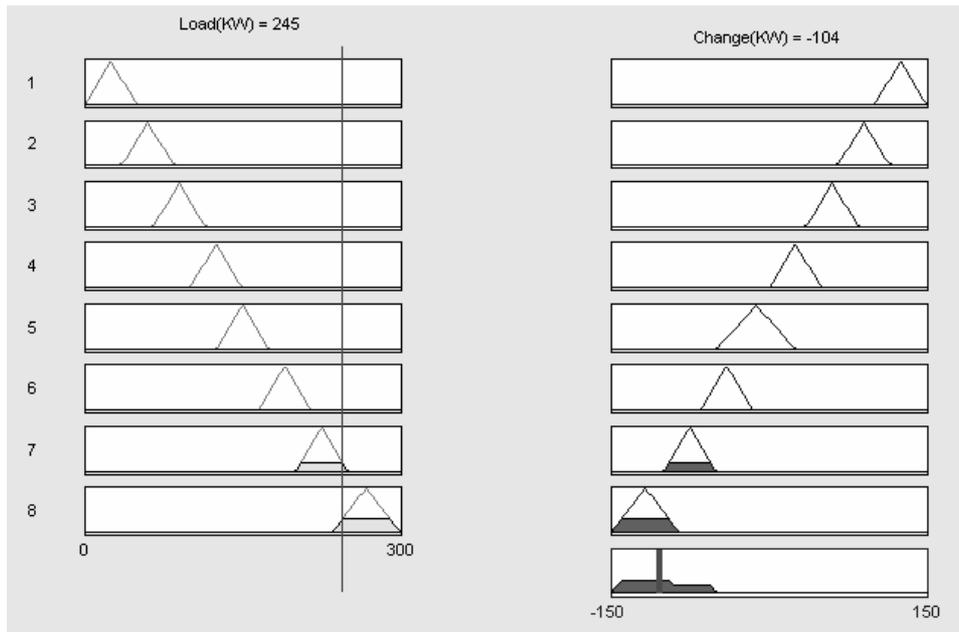

Fig. 6. Determination of the output load change for phase 1 of the input load.

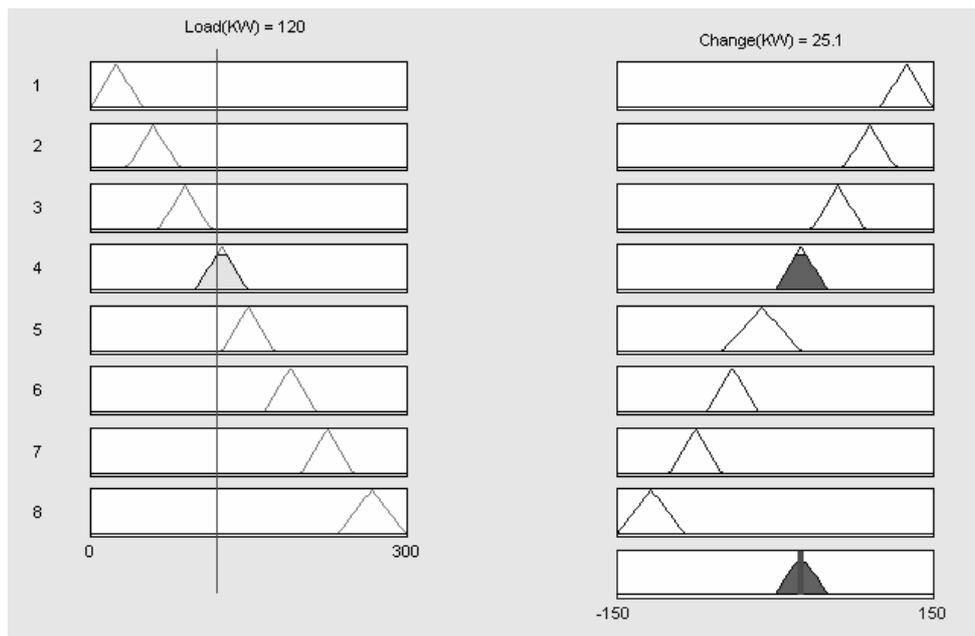

Fig. 7. Determination of the output load change for phase 2 of the input load.

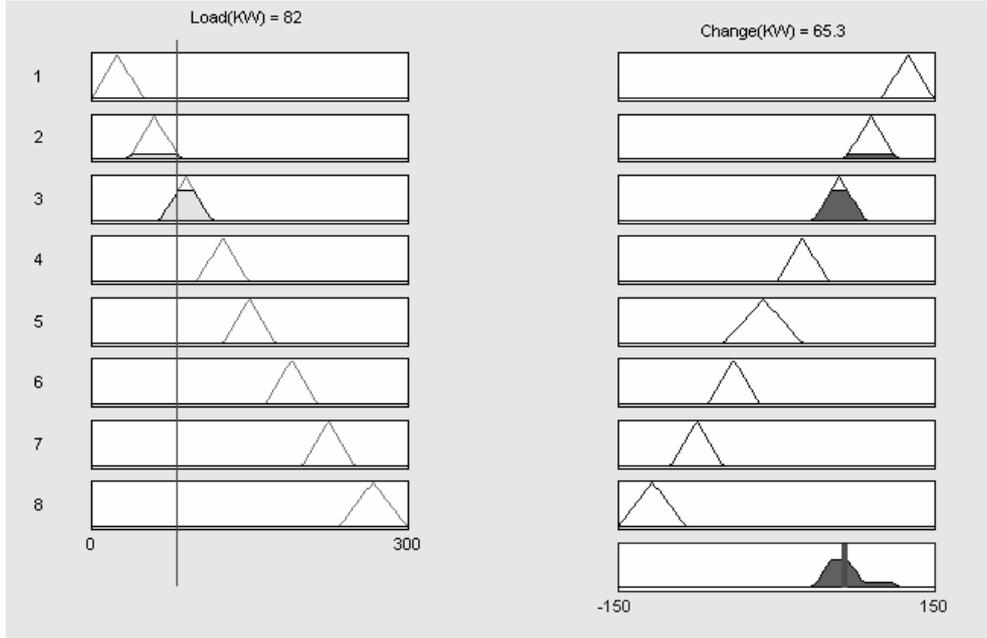

Fig. 8. Determination of the output load change for phase 3 of the input load.

So, after rounding the output load change, for the input load $P_{in} = \begin{bmatrix} 245 \\ 120 \\ 82 \end{bmatrix}$ kW, the output load change configuration is $\Delta P_{fuzzy} = \begin{bmatrix} -104 \\ 25 \\ 65 \end{bmatrix}$ kW.

However, with this load change configuration, we will have error. Because the positive and the negative totals are not equal, i.e., $\sum \Delta P_{fuzzy} = -14 \neq 0$ kW. So, if we implement this load change configuration, this will result in reduction of -14 kW of total load. This is not possible, because, with the load balancing we can only interchange the load points amongst the three phases, keeping the total load same, i.e., without increasing or decreasing the total load.

So, we have to perform an error correction. The average error (AE) is given as

$$AE = round\left(\frac{\sum \Delta P_{fuzzy}}{3}\right). \tag{3}$$

We use the average error to construct the error matrix $\Delta P_{error}$, by distributing the AE evenly among the three phases.

$$\Delta P_{error} = \begin{bmatrix} AE \\ AE \\ \sum \Delta P_{fuzzy} - 2*AE \end{bmatrix}. \tag{4}$$

We get the final load change configuration $\Delta P$, by subtracting the $\Delta P_{error}$ from the uncorrected fuzzy output $\Delta P_{fuzzy}$.

$$\Delta P = \Delta P_{fuzzy} - \Delta P_{error}, \quad \sum \Delta P = 0. \tag{5}$$

Applying (3)-(5) in our example case, we get the following.

$$AE = -5 \text{ kW}, \quad \Delta P_{error} = \begin{bmatrix} -5 \\ -5 \\ -4 \end{bmatrix} \text{kW}, \quad \Delta P = \begin{bmatrix} -104 \\ 25 \\ 65 \end{bmatrix} - \begin{bmatrix} -5 \\ -5 \\ -4 \end{bmatrix} = \begin{bmatrix} -99 \\ 30 \\ 69 \end{bmatrix} \text{kW}.$$

The final output should be, $P_{final} = P_{in} + \Delta P = \begin{bmatrix} 245 \\ 120 \\ 82 \end{bmatrix} + \begin{bmatrix} -99 \\ 30 \\ 69 \end{bmatrix} = \begin{bmatrix} 146 \\ 150 \\ 151 \end{bmatrix}$ kW.

Applying (2) on $P_{in}$ and $P_{final}$, we get respectively,

*Initial Absolute Average Unbalance* $(IAUB)$ / *Phase* $= 108.67$ kW,

*Final Absolute Average Unbalance* $(FAUB)$ / *Phase* $= 3.33$ kW.

The reduction of unbalance indicates improvement of the phase balancing.

Table 4 shows more application results for different feeder load configurations. The input loads for this study were acquired from a load data survey in a South African city [4]. The input superset consists of many load data for the three phases for any specific time-period of a day over a month. We selected the input loads randomly from the superset. The results presented in Table 4 are chosen to represent different phase loading conditions. For each application, the final fuzzy output change configuration is passed onto the implementation system for implementing the load change operation, described in the following section.

Table 4. Application results for different phase loading.

| Test Case | Initial Load (KW) | IAUB/Phase (KW) | Initial Fuzzy Change (KW) | Error (KW) | Final Fuzzy Change (KW) | Final Load (KW) | FAUB/Phase (KW) |
|---|---|---|---|---|---|---|---|
| 1 | $\begin{bmatrix}157\\134\\120\end{bmatrix}$ | 24.67 | $\begin{bmatrix}-12\\5\\25\end{bmatrix}$ | $\begin{bmatrix}6\\6\\6\end{bmatrix}$ | $\begin{bmatrix}-18\\-1\\19\end{bmatrix}$ | $\begin{bmatrix}139\\133\\139\end{bmatrix}$ | 4 |
| 2 | $\begin{bmatrix}140\\145\\156\end{bmatrix}$ | 10.67 | $\begin{bmatrix}-1\\-12\\-5\end{bmatrix}$ | $\begin{bmatrix}-6\\-6\\-7\end{bmatrix}$ | $\begin{bmatrix}5\\0\\-5\end{bmatrix}$ | $\begin{bmatrix}145\\145\\151\end{bmatrix}$ | 4 |
| 3 | $\begin{bmatrix}205\\170\\162\end{bmatrix}$ | 30 | $\begin{bmatrix}-52\\-21\\-12\end{bmatrix}$ | $\begin{bmatrix}-28\\-28\\-29\end{bmatrix}$ | $\begin{bmatrix}-24\\7\\17\end{bmatrix}$ | $\begin{bmatrix}181\\177\\179\end{bmatrix}$ | 2.67 |
| 4 | $\begin{bmatrix}170\\95\\83\end{bmatrix}$ | 58 | $\begin{bmatrix}-21\\60\\64\end{bmatrix}$ | $\begin{bmatrix}34\\34\\35\end{bmatrix}$ | $\begin{bmatrix}-55\\26\\29\end{bmatrix}$ | $\begin{bmatrix}115\\121\\112\end{bmatrix}$ | 6 |
| 5 | $\begin{bmatrix}117\\74\\42\end{bmatrix}$ | 50.67 | $\begin{bmatrix}25\\76\\108\end{bmatrix}$ | $\begin{bmatrix}70\\70\\69\end{bmatrix}$ | $\begin{bmatrix}-45\\6\\39\end{bmatrix}$ | $\begin{bmatrix}72\\80\\81\end{bmatrix}$ | 6 |

IAUB: Initial Absolute Average Unbalance, FAUB: Final Absolute Average Unbalance

## 5      Load Change Implementation System

In this section, we discuss the implementation system architecture for implementing the load changes amongst the three unbalanced phases to achieve the balance condition. Input to the implementation system, the load change values (positive and negative), comes from the fuzzy step as described in the previous section.

First, the system decides if the suggested load changes are at all possible considering the actual load values of the three phases. If it is possible, then the operation of the implementation system is divided into two steps, *determine* and *distribute*. In the determine step, the system selects the phases with negative load change suggestion, i.e., load points to be released. Considering the change values, it selects optimally the sets of load points to be shifted from the releasing phases. In the distribute step, the system considers the phases with positive (receiving) load change suggestion and distributes the load points, selected in the determine step, optimally among those receiving phases. So, the operation of the implementation system should always be a combination of determine and distribute together, i.e., the load change suggestion from the fuzzy step must have both the positive and the negative terms. If the load change suggestion is monotonically releasing (negative) or receiving (positive), the imperative load change operation is not possible. However, this should be tackled by the error correction step discussed in the previous section.

*5.1    Determine Step*

Fig. 9 shows the algorithmic flowchart of the determine step. First, for all the three phases, it is checked whether the suggested load change value (absolute) is greater than the minimum load point of that phase. If this condition fails for at least one phase, the entire load change operation is not possible. Because, in that case, the sum of the total change will not be zero, i.e., there will be some change in the total load value which is not possible as shown in (5).

Next, we determine the load points to be shifted for the releasing phases. We iterate over the three phases to check the negative (releasing) ones. For each of the releasing phases, first we calculate the average load value of that phase. Then, we determine the number of load points to be shifted from that phase by dividing the load change value by the average value and rounding the result.

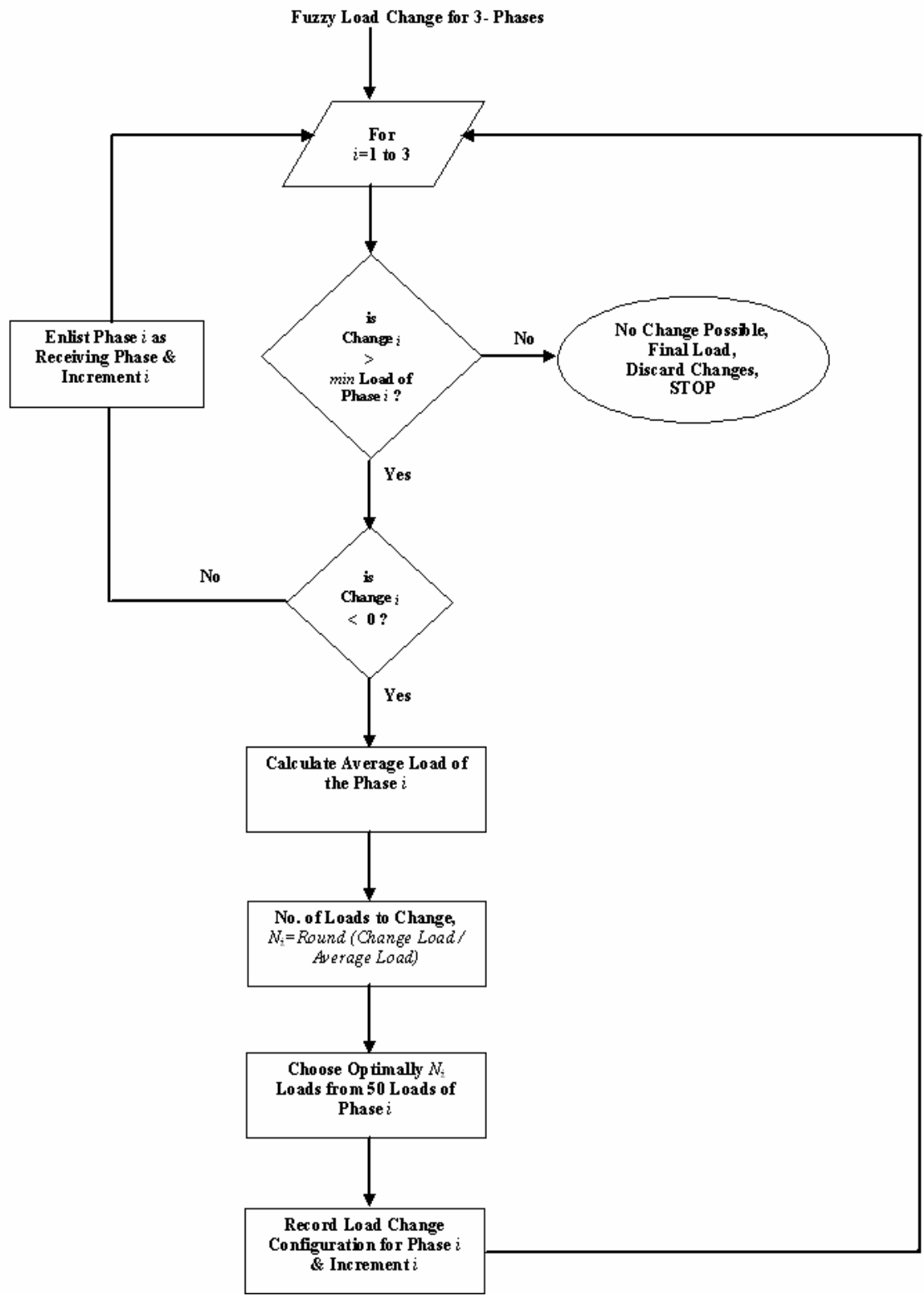

Fig. 9. Determine step of the implementation system for load change implementation.

$$\text{No. of load points to change, } N_i = \text{round}\left(\frac{\text{Fuzzy Change Value}_i}{\sum_{j=1}^{50} \text{Phase Load}_j^i \Big/ 50}\right), \quad i \in \{1,2,3\}. \tag{6}$$

Applying the extended combinatorial optimization technique [4], we determine the $N_i$ specific load points to be shifted from the $i$-th phase, where $i$ indicates the releasing phases. Once, we get the specific load points to be shifted, we record it and repeat the whole operation for the next releasing phase. The load point selection algorithm, based on the heuristic algorithm [4], is briefly explained below.

The problem is to find the set of $N_i$ load points from the 50 load points (for releasing phase $i$), sum of the set being equal to the fuzzy load change value.

$$\text{Phase}_i = \{\text{Load}_j^i, j = 1,\ldots,50\}. \tag{7}$$

$$\text{Change}_i = \{\text{Load}_k^i, k = 1,\ldots, N_i\} \quad \text{where} \quad \text{Load}_k^i \in \text{Phase}_i. \tag{8}$$

Difference between the sum of the elements of the changing load point set and the fuzzy change value for that phase must be *minimum*, ideally 0. So, we need to find the optimum set of $\text{Change}_i$, performing the following optimization task,

$$\arg\min \left|\sum \text{Change}_i - \text{Fuzzy}_i\right|, \tag{9}$$

where, $\text{Fuzzy}_i$ indicates the fuzzy load change value for the $i$-th releasing phase. Fig. 10 shows the flowchart for this selection process.

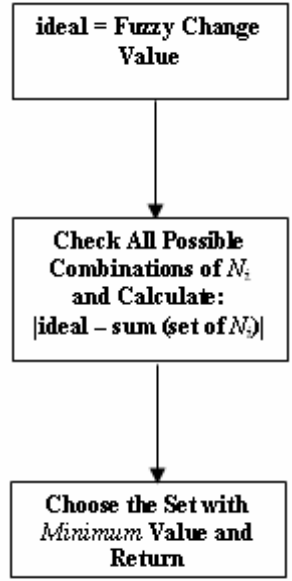

Fig. 10. Load point selection algorithm for the determine step.

*5.2    Distribute Step*

In this step, first we identify the receiving phases, i.e., the phases with positive load change value from the fuzzy step. Note that, the number of receiving phases must be 1 or 2 corresponding to the respective number of the releasing phases, 2 or 1. Neither of the two categories (releasing, receiving) can attain a value of 3. Considering this, we design our distribute step. The algorithm of the distribute step is shown as flowchart in Fig. 11.

First, we concatenate the changing load points, from the determine step, into a single change vector. Then, if there is a single receiving phase, we allocate all elements of the change vector to that phase. Otherwise, if we have two receiving phases, we have to go for optimal distribution of the load points amongst the two phases, which we describe below.

For the two receiving phase case, first we get the average of the change vector. Then we divide the fuzzy change value of one of the receiving phases with this average value to determine how many load points should be shifted to that receiving phase. Then, we go for the same combinatorial optimization based load point selection as described in section 5.1. Here, we consider the change vector as our master set and select the optimum set from that to be shifted to the one of the receiving phases in question. Once we get that, we allocate the remaining load points to the remaining receiving phase. As we work using the change vector, even if the sum of the load points transferred to one of the receiving phases does not tally with the load change value suggested by the fuzzy step, we do not alter the total load. The other receiving phase gets the rest of the released loads which makes the positive and negative terms equal even if the suggested and the implemented load changes do not match.

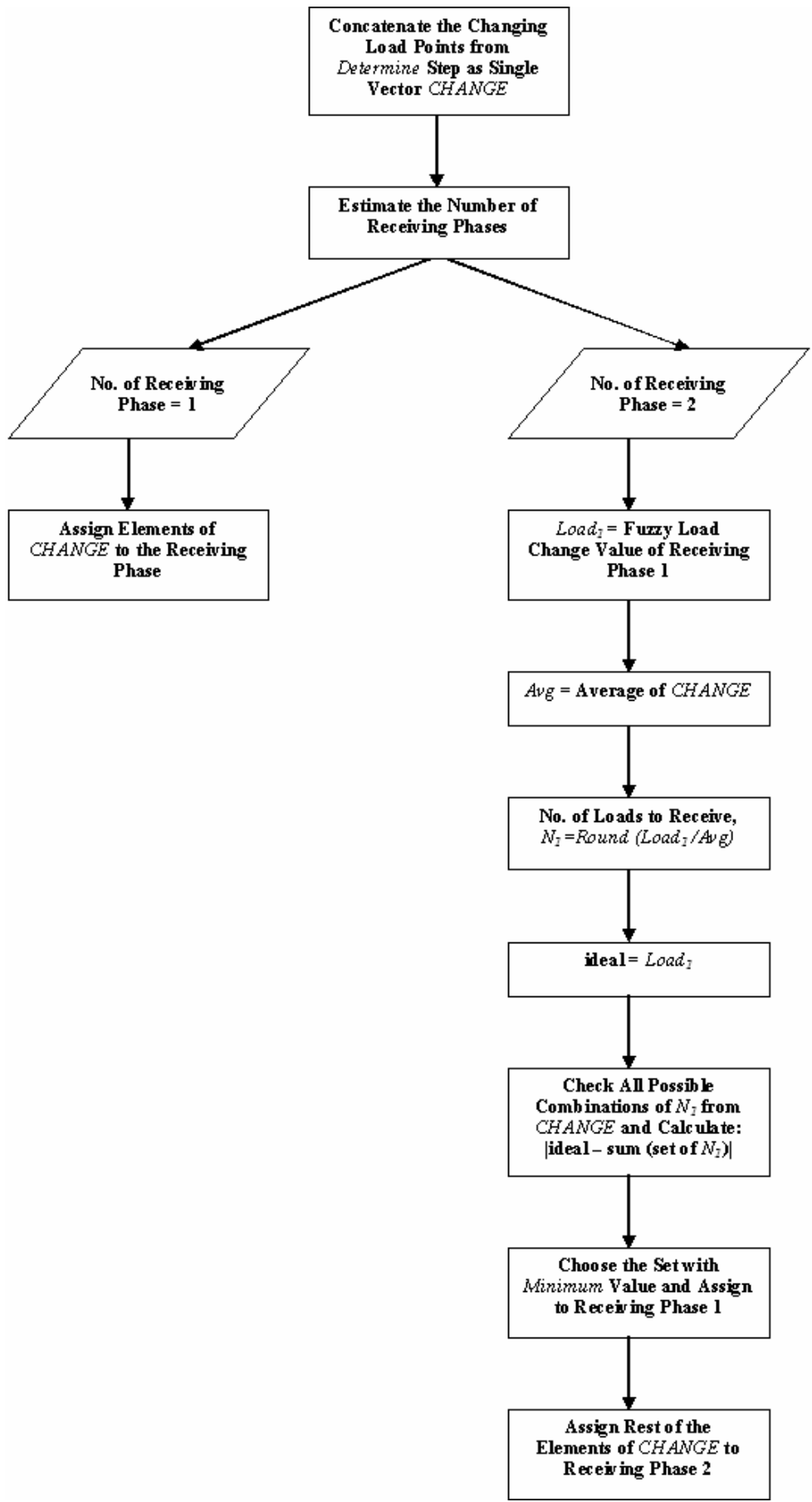

Fig. 11. Distribute step of the implementation system for load change implementation.

*5.3    Application Example*

In this section, we present the load change implementation of the application example discussed in section 4.3. We had, input load $P_{in} = [245\ 120\ 82]^T$ kW, and the fuzzy load change, $\Delta P = [-99\ 30\ 69]^T$ kW. From the fuzzy load change values, here, phase 1 is releasing and phase 2 & 3 are the receiving phases.

From the *determine* step, we get the average load change of phase 1 = 245/50 = 4.9 kW. So, the number of load points to be released from phase 1, $N_1 = round\left(\dfrac{99}{4.9}\right) = 20$.

Table 5 shows the exact load distributions of the three phases.

Table 5. Example load distribution for the three phases.

| SL. No. | Phase 1 (kW) per house | Phase 2 (kW) per house | Phase 3 (kW) per house |
|---|---|---|---|
| 1 | 5 | 4 | 1 |
| 2 | 4 | 2 | 1 |
| 3 | 4 | 3 | 2 |
| 4 | 3 | 1 | 1 |
| 5 | 2 | 3 | 3 |
| 6 | 1 | 2 | 2 |
| 7 | 2 | 1 | 1 |
| 8 | 3 | 3 | 3 |
| 9 | 5 | 1 | 1 |
| 10 | 2 | 2 | 2 |
| 11 | 5 | 5 | 2 |
| 12 | 2 | 4 | 2 |
| 13 | 6 | 2 | 3 |
| 14 | 1 | 1 | 1 |
| 15 | 2 | 4 | 2 |
| 16 | 2 | 2 | 1 |
| 17 | 5 | 3 | 2 |
| 18 | 1 | 4 | 1 |
| 19 | 2 | 2 | 2 |
| 20 | 2 | 2 | 2 |
| 21 | 3 | 1 | 1 |
| 22 | 6 | 2 | 1 |
| 23 | 8 | 3 | 2 |
| 24 | 10 | 2 | 1 |
| 25 | 9 | 1 | 1 |
| 26 | 4 | 3 | 2 |
| 27 | 2 | 4 | 1 |
| 28 | 3 | 3 | 2 |
| 29 | 5 | 2 | 1 |
| 30 | 6 | 1 | 2 |
| 31 | 6 | 2 | 3 |
| 32 | 6 | 3 | 1 |
| 33 | 9 | 2 | 2 |
| 34 | 3 | 2 | 1 |
| 35 | 10 | 2 | 1 |
| 36 | 12 | 3 | 2 |
| 37 | 15 | 4 | 1 |
| 38 | 3 | 1 | 1 |
| 39 | 2 | 2 | 3 |
| 40 | 1 | 1 | 1 |
| 41 | 3 | 2 | 2 |
| 42 | 2 | 4 | 1 |
| 43 | 1 | 3 | 2 |
| 44 | 5 | 2 | 2 |
| 45 | 4 | 1 | 1 |
| 46 | 15 | 1 | 1 |
| 47 | 12 | 2 | 2 |
| 48 | 10 | 5 | 3 |
| 49 | 9 | 2 | 1 |
| 50 | 2 | 3 | 2 |
| Total | 245 | 120 | 82 |

We apply the combinatorial optimization to select optimally 20 load points from phase 1 to get a releasing sum of 99 kW. The result from the implementation system is shown in Table 6.

Table 6. Optimal selection of 20 releasing load points from phase 1.

| Sl. No. | 1 | 2 | 3 | 4 | 5 | 6 | 7 | 8 | 9 | 10 | 11 | 12 | 13 | 14 | 15 | 16 | 17 | 18 | 19 | 20 | Total |
|---|---|---|---|---|---|---|---|---|---|---|---|---|---|---|---|---|---|---|---|---|---|
| Element No. | 1 | 5 | 9 | 14 | 15 | 16 | 21 | 22 | 23 | 24 | 25 | 26 | 27 | 28 | 29 | 30 | 31 | 32 | 33 | 44 | |
| kW Value | 5 | 2 | 5 | 1 | 2 | 2 | 3 | 6 | 8 | 10 | 9 | 4 | 2 | 3 | 5 | 6 | 6 | 6 | 9 | 5 | 99 |

In the *distribute* step, we optimally distribute these 20 load points among the receiving phases 2 and 3. Here, change vector is the set of 20 releasing load points shown in Table 6.

We get the average of the change vector= $99/20 = 4.95$ kW.

So, number of load points to be received by phase 2, $N_2 = round\left(\dfrac{30}{4.95}\right) = 6$,

and number of load points to be received by phase 3, $N_3 = 20 - 6 = 14$.

Then we use the implementation system to select optimally first the 6 loads to be transferred to the phase 2, shown in Table 7. The rest 14 loads, allocated to phase 3, are shown in Table 8.

Table 7. Optimal selection of 6 receiving load points for phase 2 from the change vector.

| Sl. No. | 1 | 2 | 3 | 4 | 5 | 6 | Total |
|---|---|---|---|---|---|---|---|
| Element No. | 1 | 2 | 7 | 8 | 9 | 16 | |
| kW Value | 5 | 2 | 3 | 6 | 8 | 6 | 30 |

Table 8. Optimal selection of 14 receiving load points for phase 3 from the change vector.

| Sl. No. | 1 | 2 | 3 | 4 | 5 | 6 | 7 | 8 | 9 | 10 | 11 | 12 | 13 | 14 | Total |
|---|---|---|---|---|---|---|---|---|---|---|---|---|---|---|---|
| Element No. | 3 | 4 | 5 | 6 | 10 | 11 | 12 | 13 | 14 | 15 | 17 | 18 | 19 | 20 | |
| kW Value | 5 | 1 | 2 | 2 | 10 | 9 | 4 | 2 | 3 | 5 | 6 | 6 | 9 | 5 | 69 |

An Intel® Celeron® 1.9 GHz, 256 MB RAM computer was used for the load change implementation simulation. The computation times were 1.51s (Matlab time) for the 20 (out of 50) load points selection and 0.27s (Matlab time) for the 6 (out of 20) load points selection.

# 6 Discussion & Future Direction

In this section, we provide the discussions of the proposed techniques and results, following up with generalizations and future directions.

## 6.1 Assumptions

In this paper, we have discussed about the phase balancing problem in the power systems domain using fuzzy logic and combinatorial optimization-based implementation system. In doing so, the following assumptions have been made for the techniques and the application examples presented in this paper.

- For the phase balancing problem we have assumed a three-phase, four-wire, 50Hz system with a radial or open loop structure, following the distribution system in South Africa.
- Each feeder is assumed to have 50 connections, i.e., 150 total connections at any point of time.
- Average domestic load of 3 kW is assumed per load per phase, i.e., 150 kW capacity per phase.
- A 200% maximum overload condition has been considered for designing the fuzzy logic-based load balancing system. Beyond this, the fuzzy logic-based load balancing system should not be used for phase balancing.
- The threshold average unbalance per phase is set at 10 kW. Over this value, phase balancing is initiated, otherwise not.
- The total load, at any point of time, remains constant, i.e., during the phase balancing, only inter-changing of the load points is possible, not increase or decrease of the total load.
- If load changing is not possible for at least one phase, the complete load changing operation is discarded even if the fuzzy step suggests some load change value. This is to keep to total load unchanged.

*6.2    Generalizations*

The above-mentioned assumptions and the techniques presented in this paper can be generically extended in the following manner.

- Although the fuzzy logic-based load balancing system is designed considering the 150 kW per phase capacity, it can be extended for any feeder capacity. This is because, for the 150 kW per phase system, we designed the system considering the maximum overload condition of 300 kW per phase. Similarly, for $X$ kW per phase system, we can generically consider the overload condition as $nX$ kW. Accordingly, we can reconcile the fuzzy rules and redesign the controller following the template shown in this paper.
- Different fuzzification and defuzzification techniques other than the Mamdani technique [17] presented in this paper, e.g., the Sugeno technique [18], can be utilized.
- The implementation system is based on the combinatorial optimization, considering a maximum of 50 load points per phase. That is, at any point of time, we are optimally choosing $n$- element set from a set of maximum 50 elements. Here, $n$ represents the number of load points to be shifted for phase balancing. However, the same optimization algorithms can be used for any amount of load points other than 50 per phase. The only difference could be the operating time, more or less depending on the number of load points.

*6.3    Future Directions*

The phase-balancing technique presented in this paper is based on the recorded distribution feeder data. Future directions extending the techniques presented here are outlined below.

- The phase balancing techniques could be applied for on-line phase balancing.
- The neural network techniques [5] and the fuzzy logic-based technique presented in this paper could be combined towards a neuro-fuzzy [19] phase balancing technique.

- The fuzzy controller and the implementation system could be implemented using field programmable gate array (FPGA) chips [20]. This is one of the major motivation of this study to utilize fuzzy logic for the phase balancing purpose.
- The proposed technique is helpful if there exists facility of phase-transfer. Therefore, extensive future task would be to investigate pilot installations of such phase-transfer facility and to apply the proposed phase balancing technique.
- Studies on power losses, voltage profiles should be undertaken with the implementation of the phase balancing system along with phase-transfer facility.

# 7    Conclusion

Phase balancing is very important and usable operation to reduce distribution feeder losses and improve system security. In this paper, we have presented a fuzzy logic-based load balancing system along with a combinatorial optimization-based implementation system for implementing the load changes. The input to the fuzzy step is the total load (kW) per phase of the feeders. Output of the fuzzy step is the load change values, negative value for load releasing and positive value for load receiving. Sum of the positive and negative values is zero, i.e., the total load remains unchanged for the entire phase balancing. The output of the fuzzy step is the input to the load changing system. The implementation system uses combinatorial optimization techniques to translate the change values (kW) into the number of load points and then selects the specific load points. It also performs the optimal inter-changing of the load points between the releasing and the receiving phases. The load balancing system is tested at the three-phase, four-wire unbalanced feeders. Application of the proposed system is substantiated by detailed application example using Matlab® for the simulations. Further application results for different feeder loading configurations indicate substantial improvement of the unbalance conditions. The proposed phase balancing system using the fuzzy logic and the implementation system is practically effective for reducing the feeder unbalance. The phase balancing techniques and the systems presented in this paper could be generically extended further for other distribution systems and feeder load configurations than presented in this paper.

**Figure Captions**

**Table Captions**